\documentclass{article}

\usepackage{authblk}

\usepackage{amsmath}
\usepackage{amssymb}
\usepackage{amsthm}

\newcommand{\sd}{\mathrm{d}}
\newcommand{\vect}[1]{\boldsymbol{#1}}

\author{D. Venhoek\footnote{{\tt david@venhoek.nl}}}
\affil{\small Radboud University Nijmegen\\
Institute for Mathematics, Astrophysics and Particle Physics}
\date{}
\title{Analyzing ``magnetic moments in curved spacetime'': pitfalls in GR}

\begin{document}
\maketitle

\begin{abstract}
We analyze the classical approximations made in ``The general relativistic effects to the magnetic moment in the Earth's gravity'', originally published as ``Post-Newtonian effects of Dirac particle in curved spacetime - I : magnetic moment in curved spacetime'', and work out precisely where in the argument the mistakes are made. We show explicitly that any difference vanishes when properly distinguishing between coordinate and physical distance. In doing this, we illustrate some of the pitfalls in using GR to make predictions.
\end{abstract}

\section{Introduction}

Morishima, Futamase and Shimizu recently published a series of papers on the influence of gravity in measuring the magnetic moment of muons, which have since been combined into~\cite{origpaper}. They predicted an order $\frac{GM}{rc^2}$ correction to the magnetic moment that could explain the discrepancy between the magnetic moment of the muon and electron, described in~\cite{PhysRevD.73.072003}. It was first noted by Visser in~\cite{firstprinciples} that these results violated core principles of general relativity, most notably the Einstein equivalence principle.

Although the first-principles arguments in~\cite{firstprinciples} are convincing, there is value in knowing where the error arose from. Nikoli\'c published a short paper~\cite{timecoord} indicating the interpretation of the time coordinate as a potential culprit, which we indeed find as part of the problem. However, Nikoli\'c left open exactly how this cancelled the correction factor found in~\cite{origpaper}.

It turns out that the misinterpretation of time on its own is not enough to account for the entire result. We will demonstrate here that the erroneous interpretation of general coordinate distances as physical distances, together with a few arithmetical errors in~\cite{origpaper} can fully account for the results found. Furthermore, we show that by correcting for these problems, a physically reasonable result is obtained.

We will not make a full analysis of the effects of gravity on measurements of the muon and electron magnetic moments here. A thorough derivation of this has been given by Laszlo and Zimboras~\cite{GRworkout}.

\section{Preliminaries}

Following along with~\cite{origpaper}, we start with the isotropic form of the Schwarzschild metric, using the mostly minus sign convention.
\begin{align}
\sd s^2 = \frac{\left(1-\frac{GM}{2rc^2}\right)^2}{\left(1+\frac{GM}{2rc^2}\right)^2}c^2\sd t^2 - \left(1+\frac{GM}{2rc^2}\right)^4\left(\sd x^2+\sd y^2+\sd z^2\right)
\end{align}

Expanding in terms of $\epsilon=\frac{1}{c}$, and substituting $\phi = -\frac{GM}{r}$, we find the expanded form
\begin{align}
\sd s^2 = \epsilon^{-2}\left(1+2\epsilon^2\phi+2\epsilon^4\phi^2\right)\sd t^2 - \left(1-2\epsilon^2\phi\right)\left(\sd x^2+\sd y^2+\sd z^2\right) + O(\epsilon^4)
\end{align}
matching the paper. Although there are legitimate doubts to be had about an expansion in terms of the speed of light (instead of for example the combination $\epsilon^2\phi$, which is dimensionless), we will not go into that here.

Instead, we will focus on a thorough derivation of the equations of motions for charged particles in such a spacetime in the presence of electromagnetic fields, trying to reconstruct the results from~\cite{origpaper}, whilst illustrating where the problems originate from.

The equations of motions are given by
\begin{align}
\frac{\sd u^\mu}{\sd\tau} + \Gamma^{\mu}_{\nu\lambda}u^\nu u^\lambda = \frac{e}{m}F^{\mu\nu}u_\nu.
\end{align}

The goal will be to transform this into a form containing $\frac{\sd\vect{\beta}}{\sd t}$ and $E$ and $B$. To do this, we need to formalize a number of definitions:

First of all, although in flat spacetime its meaning is quite clear, in curved spacetime there might be (and in fact are) multiple reasonable options to define $\beta$. To the best knowledge of the author, the definition used in~\cite{origpaper} is $\vect{\beta} = \frac{1}{c}\frac{\sd\vect{x}}{\sd t}$, where $\vect{x}$ consists of the three spatial coordinates $x$, $y$ and $z$. Rewriting, we find the useful identity $\beta = \frac{1}{c}\frac{\vect{u}}{u^0}$. Note that this is a rather simplistic definition of $\beta$, which as we will see later is part of the cause for the problems in interpreting the results.

Second, we need to relate the fields $E$ and $B$ to the electromagnetic field tensor. In the paper, this is done through vierbein fields. An equivalent approach that does not require the development of the theory of vierbein fields is to require the electromagnetic field tensor to transform in such a way that when a coordinate transformation creates a point $x^\mu$ where $g^{\mu\nu}$ is of diagonal form $(1,-1,-1,-1)$, then the electromagnetic field tensor at that point is of the standard form for flat spacetime. A short calculation then shows that for the metric above, we have (up to order $\epsilon^3$):
\begin{align}
F^{\mu\nu} = \begin{pmatrix}
0 & -\epsilon E_x & -\epsilon E_y & -\epsilon E_z\\
\epsilon E_x & 0 & -\left(1+2\epsilon^2\phi\right)B_z & \left(1+2\epsilon^2\phi\right)B_y\\
\epsilon E_y & \left(1+2\epsilon^2\phi\right)B_z & 0 & -\left(1+2\epsilon^2\phi\right)B_x\\
\epsilon E_z & -\left(1+2\epsilon^2\phi\right)B_y & \left(1+2\epsilon^2\phi\right)B_x & 0
\end{pmatrix}
\end{align}

With the preliminaries out of the way, we can now rewrite the equations of motion:
\begin{align}
\frac{\sd\vect{\beta}}{\sd t} &= \frac{\sd \tau}{\sd t}\frac{\sd}{\sd \tau}\left(\epsilon\frac{\sd \tau}{\sd t}\frac{\sd\vect{x}}{\sd\tau}\right)\nonumber\\
&= \frac{\epsilon}{u^0}\frac{\sd}{\sd \tau}\left(\frac{1}{u^0}\vect{u}\right)\nonumber\\
&= \frac{\epsilon}{\left(u^0\right)^2}\left(\frac{\sd\vect{u}}{\sd\tau} - \frac{\vect{u}}{u^0}\frac{\sd u^0}{\sd \tau}\right)
\end{align}
We now split this into coordinates, using Latin indices $i$, $j$ and $k$ to indicate indices restricted to the spatial dimensions.
\begin{align}
\frac{\sd\beta^i}{\sd t} &= \frac{\epsilon}{\left(u^0\right)^2}\left(\frac{\sd u^i}{\sd\tau} - \frac{u^i}{u^0}\frac{\sd u^0}{\sd \tau}\right)\nonumber\\
&= \frac{\epsilon}{\left(u^0\right)^2}\left(\frac{e}{m}F^{i\nu}u_\nu-\Gamma^i_{\nu\lambda}u^\nu u^\lambda-\frac{u^i}{u^0}\frac{e}{m}F^{0\nu}u_\nu+\frac{u^i}{u^0}\Gamma^0_{\nu\lambda}u^\nu u^\lambda\right)\label{eq:eomfinal}
\end{align}

We can split this equation into two parts: the gravitational contribution (all terms containing a factor $\Gamma^\mu_{\nu\lambda}$), and the electromagnetic contribution (all terms containing a factor $F^{\mu\nu}$.

\section{Gravitational contribution}

As derived above, the gravitational contribution is given by:
\begin{align}
\left(\frac{\sd\beta^i}{\sd t}\right)_{\text{grav}} &= -\frac{\epsilon}{\left(u^0\right)^2}\left(\Gamma^i_{\nu\lambda}u^\nu u^\lambda-\frac{u^i}{u^0}\Gamma^0_{\nu\lambda}u^\nu u^\lambda\right)
\end{align}

In order to rewrite this in terms of $\phi$ and $\beta$, we first need to work out the Christoffel connections. We find
\begin{align}
\Gamma^i_{00} &= \left(1+4\epsilon^2\phi\right)\frac{\sd\phi}{\sd x^i}\\
\Gamma^0_{i0} &= \epsilon^2\frac{\sd\phi}{\sd x^i}\\
\Gamma^i_{ii} &= -\epsilon^2\frac{\sd\phi}{\sd x^i}\\
\Gamma^i_{jj} &= \epsilon^2\frac{\sd\phi}{\sd x^i}\\
\Gamma^j_{ij} &= -\epsilon^2\frac{\sd\phi}{\sd x^i}
\end{align}
All other Christoffel symbols can either be found through symmetries of the Christoffel symbols themselves, or are easily seen to be zero because the metric is diagonal, symmetric in $x$, $y$ and $z$ and independent of $t$.

Using this, we can work out explicitly the $x$ component of $\frac{\sd\vect{\beta}}{\sd t}$ (the other components follow from permutation of $x$, $y$ and $z$):
\begin{align}
\left(\frac{\sd\beta^x}{\sd t}\right)_{\text{grav}} &= -\frac{\epsilon}{\left(u^0\right)^2}\left(\Gamma^x_{\nu\lambda}u^\nu u^\lambda-\frac{u^x}{u^0}\Gamma^0_{\nu\lambda}u^\nu u^\lambda\right)\nonumber\\
&= -\frac{\epsilon}{\left(u^0\right)^2}\left(\left(1+4\epsilon^2\phi\right)\frac{\sd \phi}{\sd x}\left(u^0\right)^2+\epsilon^2\frac{\sd\phi}{\sd x}\left(\left(u^x\right)^2+\left(u^y\right)^2+\left(u^z\right)^2\right)\right.\nonumber\\
&\phantom{=}\quad\quad\left.-4\epsilon^2u^x\left(\frac{\sd\phi}{\sd x}u^x + \frac{\sd\phi}{\sd y}u^y+ \frac{\sd\phi}{\sd z}u^z\right)\right)+O(\epsilon^5)
\end{align}
Working back to vector, and replacing occurrences of $\vect{u}$ with $\vect{\beta}$ then gives:
\begin{align}
\left(\frac{\sd\vect{\beta}}{\sd t}\right)_{\text{grav}} &= -\frac{\epsilon}{\left(u^0\right)^2}\left(\left(\left(1+4\epsilon^2\phi\right)\left(u^0\right)^2+\epsilon^2\vect{u}^2\right)\vect{\nabla}\phi-4\epsilon^2\vect{u}\vect{\nabla}\phi \cdot \vect{u}\right)+O(\epsilon^5)\nonumber\\
&= -\epsilon\left(\left(1+\vect{\beta}^2\right)\vect{\nabla}\phi-4\vect{\beta}\vect{\nabla}\phi\cdot\vect{\beta}\right)+O(\epsilon^3)
\end{align}

This mostly matches the result from~\cite{origpaper}, giving confidence that the definition of $\vect{\beta}$ is the same. The author suspects that the difference in powers of $\epsilon$ noted as missing in the expansion is a small mistake in~\cite{origpaper}.

\section{Electromagnetic contribution}

Let us now start rewriting the electromagnetic contribution found in Equation~\ref{eq:eomfinal}:
\begin{align}
\left(\frac{\sd\beta^i}{\sd t}\right)_{\text{em}} &= \frac{\epsilon}{\left(u^0\right)^2}\left(\frac{e}{m}F^{i\nu}u_\nu-\frac{u^i}{u^0}\frac{e}{m}F^{0\nu}u_\nu\right)
\end{align}

We start by working out the $x$ component explicitly (the other components follow from cyclic permutation of $x$, $y$ and $z$):
\begin{align}
\left(\frac{\sd\beta^x}{\sd t}\right)_{\text{em}} &= \frac{\epsilon}{\left(u^0\right)^2}\left(\frac{e}{m}F^{x\nu}u_\nu-\frac{u^x}{u^0}\frac{e}{m}F^{0\nu}u_\nu\right)\nonumber\\
&= \frac{\epsilon}{\left(u^0\right)^2}\left(\frac{e}{m}F^{x\nu}g_{\nu\lambda}u^\lambda-\frac{u^x}{u^0}\frac{e}{m}F^{0\nu}g_{\nu\lambda}u^\lambda\right)\nonumber\\
&= \frac{\epsilon}{\left(u^0\right)^2}\left(\vphantom{\frac{u^x}{u^0}}\frac{e}{m}\epsilon E_x\epsilon^{-2}\left(1+2\epsilon^2\phi+2\epsilon^4\phi^2\right)u^0\right.\nonumber\\
&\phantom{=}\quad\quad\left.+\vphantom{\frac{u^x}{u^0}}\frac{e}{m}\left(1-2\epsilon^2\phi\right)\left(1+2\epsilon^2\phi\right)\left(B_zu^y-B_yu^z\right)\right.\nonumber\\
&\phantom{=}\quad\quad\left.-\frac{u^x}{u^0}\frac{e}{m}\left(1-2\epsilon^2\phi\right)\epsilon\left(E_xu^x+E_yu^y+E_zu^z\right)\right) + O(\epsilon^5)\nonumber\\
&= \frac{1}{u^0}\frac{e}{m}\left(\left(1+2\epsilon^2\phi\right)E_x+B_z\beta^y - B_y\beta^z\right.\nonumber\\
&\phantom{=}\quad\quad\left.-\left(1-2\epsilon^2\phi\right)\beta^x\left(E_x\beta^x+E_y\beta^y+E_z\beta^z\right)\right)+O(\epsilon^3)
\end{align}
This can be rewritten in vector form, yielding:
\begin{align}
\left(\frac{\sd\vect{\beta}}{\sd t}\right)_{\text{em}} &= \frac{1}{u^0}\frac{e}{m}\left(\left(1+2\epsilon^2\phi\right)\vect{E} + \vect{\beta}\times\vect{B} - \left(1-2\epsilon^2\phi\right)\vect{\beta}\vect{E}\cdot\vect{\beta}\right)
\end{align}

Finally, we need to work out $\frac{1}{u^0}$ in terms of $\beta$ and $\gamma = \frac{1}{\sqrt{1-\beta^2}}$. Starting with $u^0$, we have by definition of $u^\mu$ $g_{\mu\nu}u^\mu u^\nu = c^2$. Rewriting this:
\begin{align}
c^2\left(1+2\epsilon^2\phi+2\epsilon^4\phi^2\right)\left(u^0\right)^2 - \left(1-2\epsilon^2\phi\right)\left(\vect{u}^2\right)&= c^2\\
c^2\left(1+2\epsilon^2\phi+2\epsilon^4\phi^2\right)\left(u^0\right)^2 - \left(1-2\epsilon^2\phi\right)c^2\beta^2\left(u^0\right)^2 &= c^2\\
\left(1+2\epsilon^2\phi+2\epsilon^4\phi^2 - \left(1-2\epsilon^2\phi\right)\beta^2\right)\left(u^0\right)^2 &= 1
\end{align}
Solving for $u^0$ now gives:
\begin{align}
u^0 &= \frac{1}{\sqrt{1+2\epsilon^2\phi+2\epsilon^4\phi^2 - \left(1-2\epsilon^2\phi\right)\beta^2}}\\
\frac{1}{u^0} &= \sqrt{1+2\epsilon^2\phi+2\epsilon^4\phi^2 - \left(1-2\epsilon^2\phi\right)\beta^2}\nonumber\\
&= \sqrt{1-\beta^2} + \epsilon^2\phi\frac{1+\beta^2}{\sqrt{1-\beta^2}}+O(\epsilon^4)\nonumber\\
&= \frac{1}{\gamma}\left(1+\epsilon^2\phi\left(2\gamma^2-1\right)\right)+O(\epsilon^4)
\end{align}

Filling this in into the electromagnetic contribution gives:
\begin{align}
\left(\frac{\sd\vect{\beta}}{\sd t}\right)_{\text{em}} &= \left(1+\epsilon^2\phi\left(2\gamma^2-1\right)\right)\frac{e}{\gamma m}\nonumber\\
&\phantom{=}\quad\quad\left(\left(1+2\epsilon^2\phi\right)\vect{E} + \vect{\beta}\times\vect{B} - \left(1-2\epsilon^2\phi\right)\vect{\beta}\vect{E}\cdot\vect{\beta}\right)
\end{align}

We note a few key differences here in comparison to the result found in~\cite{origpaper}: First of all, in the leading factor, we find a sign difference. Furthermore, note that we find rather different leading factors for the various electromagnetic field contributions. As we will see soon, these factors play a crucial role in explaining why this result is expected. The author is confident both differences result from calculation errors made in preparing~\cite{origpaper}.

\section{Interpretation of results}\label{sec:interp}

In order to interpret the result above in connection to the magnetic moment of a muon, the authors of~\cite{origpaper} consider a situation of a small-velocity particle moving in a pure magnetic field $(\vect{E} = 0)$. Applying this to our results, and ignoring the gravitational contributions, we find
\begin{align}
\frac{\sd\vect{\beta}}{\sd t} \approx \left(1+\epsilon^2\phi\right)\frac{e}{m}\beta\times B.
\end{align}

It is tempting to read this as a gravitational correction to the electromagnetic interaction between the magnetic field and the particle. However, this is not correct. Implicitly, such an interpretation interprets $\vect{\beta}$ as a velocity in terms of a fraction of the speed of light. However, given the definition of $\beta$ as change in $\vect{x}$ over change in $t$, divided by the speed of light, this fails. Since there is a factor $\left(1+2\epsilon^2\phi+2\epsilon^4\phi^2\right)$ in front of $\sd t^2$ in the metric, a unit change in $t$ no longer corresponds to a time period of $1$ second, but instead to a time period of $1+\epsilon^2\phi + O(\epsilon^4)$ seconds. Similarly, a unit change in $\vect{x}$ corresponds to a distance change of $1-\epsilon^2\phi$ meters. When interpreting $\beta$ as a velocity in terms of a fraction of the speed of light, we make the mistake of interpreting coordinate distances as physical distances, without regard for unit conversions.

In order to interpret $\beta$ as a velocity in terms of fraction of the speed of light, these effects need to be compensated for. To see how this works, let us define $\beta_{\text{ph}}$ and $\gamma_{\text{ph}}$ such that we correct for the difference between coordinates and physical units:
\begin{align}
\beta_{\text{ph}} &= \frac{1-\epsilon^2\phi}{1+\epsilon^2\phi} \beta \approx \left(1-2\epsilon^2\phi\right) \beta\\
\gamma_{\text{ph}} &= \frac{1}{\sqrt{1-\beta_{\text{ph}}^2}}
\end{align}

Let us start by rewriting the factor in front:
\begin{align}
\left(1+\epsilon^2\phi\left(2\gamma^2-1\right)\right)\frac{1}{\gamma} &= \sqrt{1-\beta^2} + \epsilon^2\phi\frac{1+\beta^2}{\sqrt{1-\beta^2}}+O(\epsilon^4)\nonumber\\
&= \sqrt{1-\left(1+4\epsilon^2\phi\right)\beta_{\text{ph}}^2}+\epsilon^2\phi\frac{1+\left(1+4\epsilon^2\phi\right)\beta_{\text{ph}}^2}{\sqrt{1-\left(1+4\epsilon^2\phi\right)\beta_{\text{ph}}^2}}+O(\epsilon^4)\nonumber\\
&= \sqrt{1-\beta_{\text{ph}}^2} -2\epsilon^2\phi\frac{\beta_{\text{ph}}^2}{\sqrt{1-\vect{\beta}_{\text{ph}}^2}}+\epsilon^2\phi\frac{1+\beta_{\text{ph}}^2}{\sqrt{1-\vect{\beta}_{\text{ph}}^2}}+O(\epsilon^4)\nonumber\\
&= \sqrt{1-\vect{\beta}_{\text{ph}}^2}\left(1+\epsilon^2\phi\right)+O(\epsilon^4)\nonumber\\
&= \frac{1}{\gamma_{\text{ph}}}\left(1+\epsilon^2\phi\right)+O(\epsilon^4)
\end{align}

Using this, we calculate $(1-\epsilon^2\phi)\left(\frac{\sd\vect{\beta}_{\text{ph}}}{\sd t}\right)_{\text{em}}$, which is the rate of change of $\beta_{\text{ph}}$ per second due to electromagnetic forces.

\begin{align}
(1-\epsilon^2\phi)\left(\frac{\sd\vect{\beta}_{\text{ph}}}{\sd t}\right)_{\text{em}} &=  (1-\epsilon^2\phi)\frac{\sd}{\sd t}\left(\left(1-2\epsilon^2\phi\right)\vect{\beta}\right)\nonumber\\
&= \left(1-3\epsilon^2\phi\right)\frac{\sd\vect{\beta}}{\sd t}\nonumber\\
&= \left(1-2\epsilon^2\phi\right)\frac{e}{\gamma_{\text{ph}} m}\nonumber\\
&\phantom{=}\quad\quad\left(\left(1+2\epsilon^2\phi\right)\vect{E} + \vect{\beta}\times\vect{B} - \left(1-2\epsilon^2\phi\right)\vect{\beta}\vect{E}\cdot\vect{\beta}\right)\nonumber\\
&= \frac{e}{\gamma_{\text{ph}} m}\left(\vect{E}+\vect{\beta}_{\text{ph}}\times\vect{B} -\vect{\beta}_{\text{ph}}\vect{E}\cdot\vect{\beta}_{\text{ph}}\right)
\end{align}

Note that this result matches that for a flat spacetime. This is expected, because derivatives of the metric do not occur in the electromagnetic part of the equations of motion. Hence, by choosing a coordinate transformation that changes the metric to a Minkowski form at a point, it follows immediately that at that point the equations of motion from electromagnetism take the form from special relativity. The fact that we could derive this result serves as a check on our results.

As a result of this, we see that general relativity does \emph{not} give rise to a modification to the local interaction between particles and electromagnetic fields. Note that this does not mean that there is no effect of gravity in experiments measuring the muon magnetic moment, but rather that the derivation for those effects takes other things into account. This is worked out in detail in~\cite{GRworkout}.

\section{Conclusion}

We have shown that, in contrast to the claims in~\cite{origpaper}, there is no change in the interaction between electromagnetic fields and a charged particle in curved spacetimes. Our results shows that any apparent difference is entirely caused by interpreting coordinate distances as physical distances. 

Furthermore, we have shown that innocent looking definitions, such as that of $\beta$, can hide significant problems when used directly in the context of general relativity. It is therefore advisable to be careful when using such constructs, and explicitly define them when necessary.

With some care, one could construct a similar argument on why interpreting changes to the quantum Hamiltonian in curved spacetime is flawed. Furthermore, one needs to be careful with attaching a physical interpretation to the quantum Hamiltonian in curved spacetimes in general, as it is not always uniquely defined~\cite{birrell1984quantum}. This makes it likely (although not proven) that the interpretation of the quantummechanical derivations in~\cite{origpaper} will also not hold up.

\section{Acknowledgements}
The author would like to thank R. Kleiss, C. Timmermans, W. Beenakker and S. Caron for the discussions that made this work possible.

\bibliography{muongravity}{}
\bibliographystyle{hunsrt}

\end{document}